\documentclass[showpacs,preprintnumbers,eqsecnum,nofootinbib,prd]{revtex4}  

\usepackage{graphicx}
\usepackage{amssymb}
\usepackage{subfigure}

\def\beq{\begin{equation}}
\def\eeq{\end{equation}}

\begin{document}

\title{Late-time evolution of cosmological models with fluids obeying a Shan-Chen-like equation of state}

\author{Donato Bini${}^{1,2}$, Giampiero Esposito$^{2}$, Andrea Geralico$^1$
} 
  \affiliation{
${}^{1}$Istituto per le Applicazioni del Calcolo ``M. Picone,'' CNR, I-00185 Rome, Italy\\
${}^{2}$Istituto Nazionale di Fisica Nucleare, Sezione di Napoli, Complesso Universitario
di Monte S. Angelo, Via Cintia, Edificio 6, 80126 Napoli, Italy}

\date{\today}

\begin{abstract}
Classical as well as quantum features of the late-time evolution of cosmological models with fluids obeying a Shan-Chen-like equation of state are studied.
The latter is of the type $p=w_{\rm eff}(\rho)\,\rho$, and has been used in previous works to describe, e.g., a possible scenario for the growth of the dark-energy content of the present Universe. 
At the classical level the fluid dynamics in a spatially flat Friedmann-Robertson-Walker background implies the existence of 
two possible equilibrium solutions depending on the model parameters, associated with (asymptotic) finite pressure and energy 
density. We show that no future cosmological singularity is developed during the evolution for this specific model.
The corresponding quantum effects in the late-time behavior of the system are also investigated within the framework of quantum geometrodynamics, i.e., by solving 
the (minisuperspace) Wheeler-DeWitt equation in the Born-Oppenheimer approximation, constructing wave-packets and analyzing their behavior.
\end{abstract}


\maketitle

\section{Introduction}

Most investigations in modern cosmology assume our Universe to be described by a Friedmann-Robertson-Walker 
(FRW) background filled by a cosmological fluid endowed with suitable properties, or equivalently 
a scalar field with appropriate (self-interacting) kinetic and potential energies.
The scalar-field dynamics should mimic different kinds of matter-energy during each phase of the cosmological 
evolution up to the present epoch, which is characterized by an accelerating expansion of the Universe. 
In this respect, we have recently investigated the role of cosmological fluids obeying an equation of 
state of the type $p=w_{\rm eff}(\rho)\,\rho$, extensively used in the context of lattice kinetic theory to model dynamic phase transitions, i.e., the 
Shan-Chen (SC) equation of state \cite{shanchen}. Its main property is to support phase transitions, so that in a cosmological context different epochs 
in the evolution of the Universe can be described as a natural consequence of the SC fluid dynamical equations.
Furthermore, because of its flexibility, it makes it possible to represent a broad variety of cosmological fluids, from radiation 
to \lq\lq exotic'' fluids with negative pressure, through a smooth variation of its free parameters.

SC-based cosmological models have been successfully applied to account for the current distribution of dark energy as well as to 
describe the inflationary epoch \cite{scprd,scinfla}.
Constraints on the free parameters of the model have been derived in Ref. \cite{sc_cmb} according to current observational data of the cosmic microwave background and baryon acoustic oscillation spectra, which enforce the validity of these models. In fact, they were previously tested only on available data from measurements of distant type Ia supernovae.
The aim of the present work is to study the late-time evolution of such models, 
hence extending our previous analysis. 
We will investigate the properties of (equilibrium) attractor solutions, including their stability against small perturbations, as well as the possible 
occurrence of future singularities. We will also develop a quantum analysis based on the 
Wheeler-DeWitt (WDW) equation \cite{wdw} in the 
framework of quantum geometrodynamics, in order to investigate the role of the associated quantum effects for large values of the scale factor.
This approach is largely used in the literature to study the quantum avoidance of classical singularities as well 
as quantum effects at large cosmological scales \cite{kiefer1,kiefer2,kamen,kiefer3}.
Other approaches include path-integral quantization, loop quantum cosmology, and string theory 
(see, e.g., Ref. \cite{kieferqg} for a recent review).

The paper is organized as follows. Section II briefly recalls the main properties of SC cosmological models and their equivalent formulation in terms of an ordinary canonical scalar field, describing then both scalar-field dynamics and late-time evolution.
The asymptotic dynamics determines the profile of the interaction potential.
Section III studies the quantum-cosmology counterpart, when the WDW equation can be reduced to a
partial differential equation, and the wave function can be expanded in terms of \lq\lq fast'' and
\lq\lq slow'' degrees of freedom according to the Born-Oppenheimer (BO) approximation. 
Explicit forms of such degrees of freedom are found for both SC attractor solutions and wave packets are constructed. 
Concluding remarks and open problems are summarized in Sec. IV.

\section{Scalar field Shan-Chen model}

Let us consider a spatially flat FRW universe, i.e., with squared line element 
${\rm d}s^2=-{\rm d}t^2+a^2({\rm d}x^2+{\rm d}y^2+{\rm d}z^2)$, 
where $a=a(t)$ is the scale factor and $c=1$, filled by a perfect fluid obeying a 
nonideal SC-like equation of state, that we write in the form
\begin{eqnarray}
\label{pscdef}
p&=&w_{\rm (in)}\rho_{\rm (crit),0}  \left[\frac{\rho}{\rho_{\rm (crit),0} }+\frac{g}{2} \psi^2\right]\,,\nonumber\\
\psi&=& 1-{\rm e}^{-\alpha_{\rm sc} \frac{\rho}{\rho_{\rm (crit),0}}},
\end{eqnarray}
where $\rho_{\rm (crit),0}=3H_0^2/\kappa^2$ is the present value of the critical density 
($H_0$ denoting the Hubble constant and $\kappa^2=8\pi G$).
The dimensionless quantities $w_{\rm (in)}$, $g$ and $\alpha_{\rm sc}$ 
are free parameters of the model, each of them carrying a 
well-defined physical meaning: $w_{\rm (in)}$ describes the nature of matter, ordinary (positive) or exotic 
(negative), in both high-density and low-density limits; $g\le 0$ measures the strength of nonideal interactions within the fluid; $\alpha_{\rm sc}\ge0$ sets 
the ratio between the actual critical density and a threshold density above which the excess pressure (with respect to the ideal gas behavior) saturates to a constant 
value, a regime often associated with \lq\lq asymptotic freedom,'' as it corresponds to a vanishing contribution of 
nonideal forces to the momentum budget of the fluid.
The idea of asymptotic freedom at high-density short-distance regimes has played a crucial role in the 
development of quantum field theory, specifically in the framework of quantum chromodynamics \citep{mandlshaw}. 

\begin{figure}
\begin{center}
\includegraphics[scale=0.35]{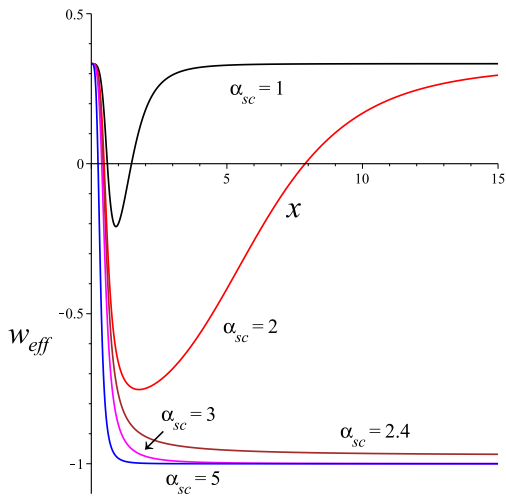}
\end{center}
\caption{
The behavior of the effective equation of state $w_{\rm eff}\equiv{p}/{\rho}$ as a function of the dimensionless 
scale factor $x$ is shown for a SC model with parameters  $w_{\rm (in)}=1/3$, $g=-8$ and different values of $\alpha_{\rm sc}=[1,2,2.4,3,5]$. 
Initial conditions are chosen so that at the present time $\tau=\tau_0$ one has $x(\tau_0)=1$ and $\xi(\tau_0)=1$.
Notice that for large values of the scale factor $w_{\rm eff}\to w_{\rm (in)}$ for $\alpha_{\rm sc}=[1,2,2.4]$, whereas $w_{\rm eff}\to -1$ for $\alpha_{\rm sc}=[3,5]$, showing the existence of a critical value of $\alpha_{\rm sc}$ ($\alpha_{\rm sc}^{\rm crit}\approx2.455$), as discussed in the text.
For $x\to0$, instead, $w_{\rm eff}$ approaches the common value $w_{\rm (in)}=1/3$.
}
\label{fig:weff}
\end{figure}

Under the assumption that the SC fluid dominates over other matter-energy degrees of freedom, the field equations are 
\begin{equation}
\label{FRWeqs}
H^2=\frac{\kappa^2}{3}\rho,\qquad
\dot\rho=-3H(\rho +p),
\end{equation}
where $H=\dot a/a$ is the Hubble parameter, and an overdot indicates derivative with respect to the cosmic time $t$.
It is convenient to introduce the following set of dimensionless variables:
\begin{equation}
\label{adimvar}
\xi \equiv \frac{\rho}{\rho_{\rm (crit),0}}, \qquad
x \equiv \frac{a}{a_0}, \qquad 
\tau \equiv H_0t,
\end{equation}
so that Eqs. (\ref{FRWeqs}) become
\begin{equation}
\label{FRWeqs2}
\frac{{\rm d}x}{{\rm d}\tau }=x\sqrt{\xi},\qquad
\frac{{\rm d}\xi}{{\rm d}\tau }=-3\sqrt{\xi}[\xi +w_{\rm (in)}{\mathcal P}(\xi)],
\end{equation}
where
\begin{equation}
\label{mathP}
{\mathcal P}(\xi)=\xi +\frac12 g \psi^2(\xi), \qquad
\psi(\xi)=1-{\rm e}^{-\alpha_{\rm sc} \xi }.
\end{equation}
The (rescaled, dimensionless) pressure ${\mathcal P}(\xi)$ reduces to the ideal gas expression ${\mathcal P}(\xi) \sim \xi$  in the low-density limit, whereas in the high-density limit it gives
${\mathcal P} \sim \xi + \frac 12 g$, i.e., the excess pressure is just a constant, often associated with vacuum fluctuations.
Noticeably,  by changing the parameters of the model, i.e., $w_{\rm (in)}$, $g$ and $\alpha_{\rm sc}$, 
the SC equation of state can attain a broad range of values of cosmological interest for the ratio 
\begin{equation}
\label{weffdef}
w_{\rm eff}\equiv \frac{p}{\rho}
=w_{\rm(in)}\left(1+\frac{g}{2} \frac{\psi^2}{\xi}\right).
\end{equation}
In fact, because of the Einstein's equations fluid dynamics, the SC fluid might undergo a transition from ordinary matter ($w_{\rm eff}>0$) to exotic matter ($w_{\rm eff}<0$) and viceversa, as shown in Fig. \ref{fig:weff}.
We refer to Ref. \cite{scprd} for more details on SC thermodynamics and the microscopic foundations of the SC equation of state.


\subsection{Late-time evolution}

\begin{figure}
\centering
\subfigure[]{\includegraphics[scale=0.35]{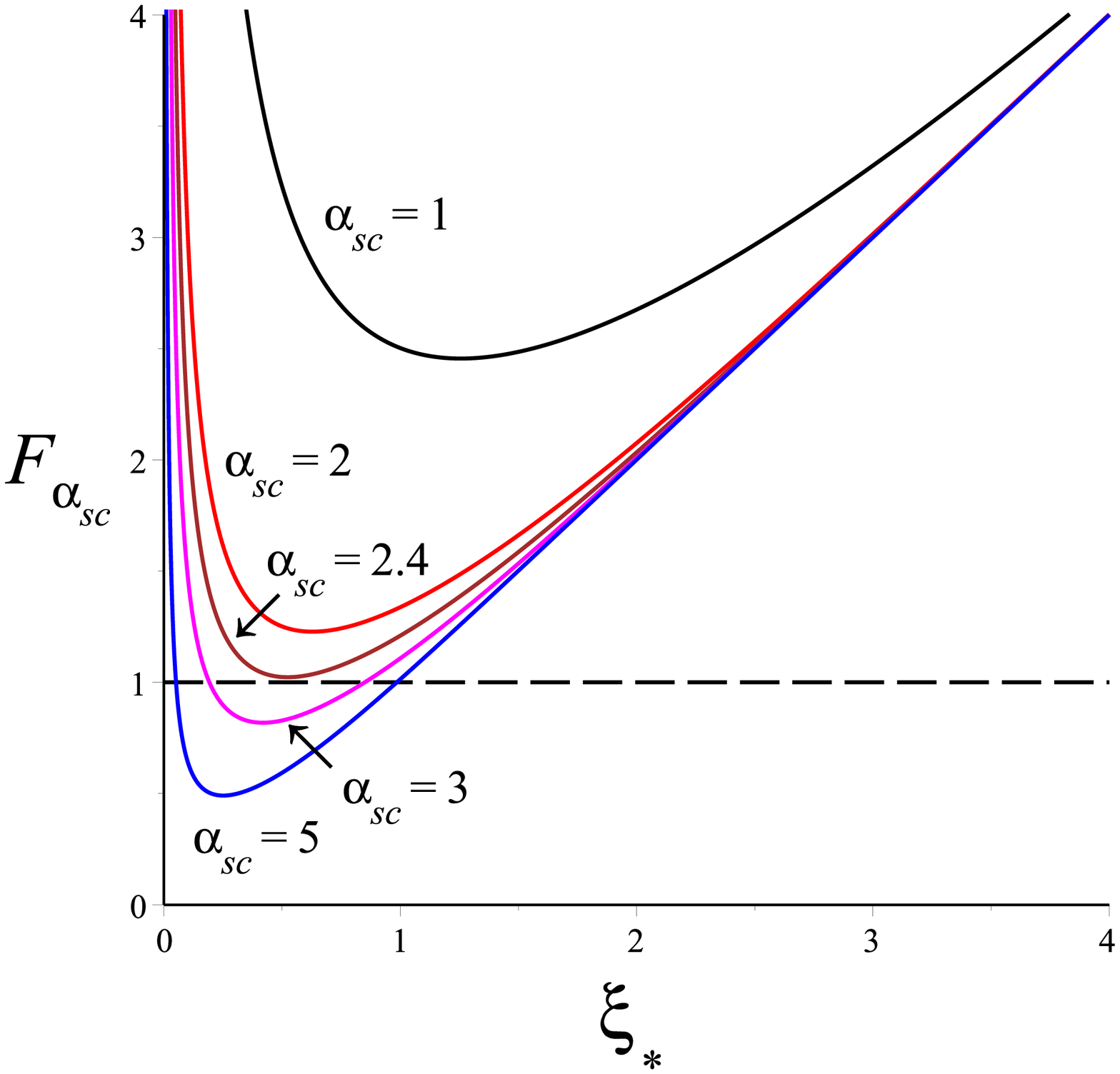}}
\hspace{5mm}
\subfigure[]{\includegraphics[scale=0.35]{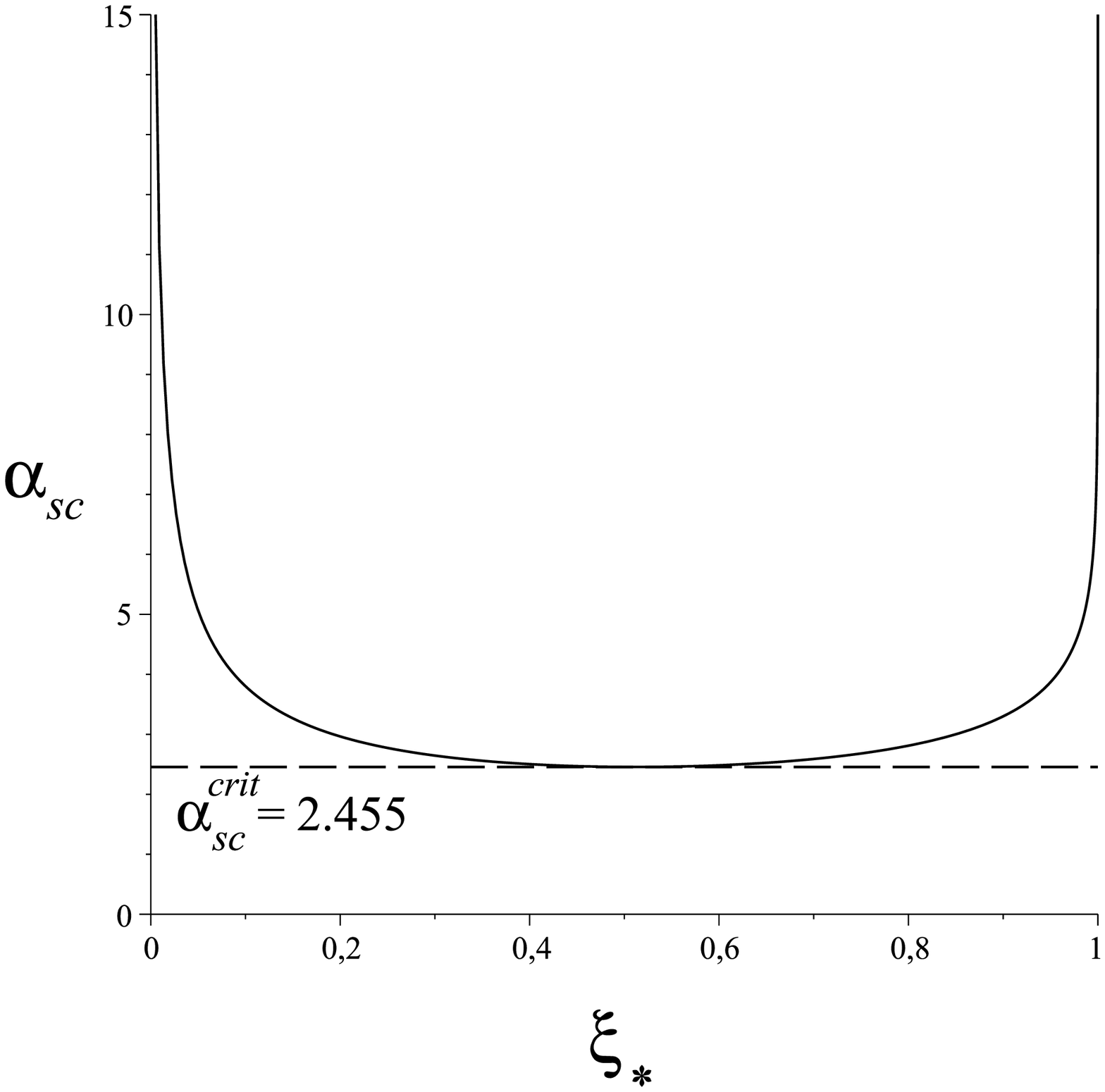}}
\caption{
Panel (a).
The behavior of the function $F_{\alpha_{\rm sc}}$ (see Eq. (\ref{eq_F})) determining the attractor solutions 
$\xi=\xi_*$ (see Eq. (\ref{eq_F})) is shown for different values of $\alpha_{\rm sc}=[1,2,2.4,3,5]$. 
The existence of two solutions is evident by drawing horizontal lines. 
The dashed line corresponds to the choice of parameters $w_{\rm(in)}=1/3$, $g=-8$, leading to $F_{\alpha_{\rm sc}}=1$, 
so that two roots do exist for $\alpha_{\rm sc}\gtrapprox2.455$.
Panel (b). 
The set of equilibrium solutions $\xi=\xi_*$ is shown as a function of $\alpha_{\rm sc}$ greater than the critical value $\alpha_{\rm sc}^{\rm crit}\approx2.455$
(dashed horizontal line). The solutions $\xi_*\to0,1$ are approached in the limit $\alpha_{\rm sc}\to \infty$.
}
\label{fig:F}
\end{figure}

We are interested here in investigating the late-time evolution 
of SC cosmological models.
The field equations (\ref{FRWeqs2}) imply that there exist attractor (equilibrium) solutions corresponding to 
$\xi=0$ and $\xi=\xi_*\not=0$ given by the roots of the equation
\begin{equation}
\label{attractor}
\xi_* +w_{\rm (in)}{\mathcal P}(\xi_*)=0,
\end{equation}
which can be at most two, for fixed values of the parameters, as already discussed in Ref. \cite{scprd}.
The previous equation, in fact, yields the condition 
\begin{equation}
\label{eq_F}
F_{\alpha_{\rm sc}}(\xi_*)\equiv \frac{\xi_*}{\left(1-{\rm e}^{-\alpha_{\rm sc} \xi_*}  \right)^2} 
= \frac{w_{\rm (in)}|g|}{2(1+w_{\rm (in)})},
\end{equation}
for fixed values of $\alpha_{\rm sc}$ and $w_{\rm (in)}>0$.
We will not consider below the case of a scalar field which behaves as a phantom field in both low and high density regimes, i.e., $w_{\rm (in)}<-1$.
If $-1\leq w_{\rm (in)}\leq0$, instead, the condition (\ref{attractor}) cannot be satisfied, so that there exists the attractor solution $\xi=0$ only. 
The function $F_{\alpha_{\rm sc}}(\xi)$ exhibits a minimum at $\xi_*^{\rm min}=-\frac{1}{2\alpha_{\rm sc}}
\left[2W_{-1}\left(-\frac{1}{2\sqrt{e}} \right)+1\right]\approx1.256/\alpha_{\rm sc}$, with value 
$F_{\alpha_{\rm sc}}(\xi_*^{\rm min})\approx2.455/\alpha_{\rm sc}$, where $W_{-1}(z)$ denotes the branch 
of the Lambert $W$ function satisfying $W(z)\leq-1$.
The existence of two roots for Eq. (\ref{eq_F}) is thus guaranteed if $2.455/\alpha_{\rm sc}\lessapprox{w_{\rm (in)}|g|}/{2(1+w_{\rm (in)})}$. 
Figure \ref{fig:F} shows the pair of attractor solutions $\xi=\xi_*$ for a selected parameter choice, 
only one of them being reached during the evolution, depending on the chosen initial conditions.
Therefore, for fixed values of the parameters $w_{\rm (in)}$ and $g$, there exists a critical value of $\alpha_{\rm sc}$ 
which discriminates between the two possible asymptotic states, i.e.,
\begin{equation}
\alpha_{\rm sc}^{\rm crit}\approx2.455\frac{2(1+w_{\rm (in)})}{w_{\rm (in)}|g|},
\end{equation}
such that $\xi\to0$ for $\alpha_{\rm sc}<\alpha_{\rm sc}^{\rm crit}$ and $\xi\to\xi_*$ for $\alpha_{\rm sc}>\alpha_{\rm sc}^{\rm crit}$.
For instance, for the choice of parameters $w_{\rm(in)}=1/3$ and $g=-8$ we get $\alpha_{\rm sc}^{\rm crit}\approx2.455$ (see Fig. \ref{fig:F}). 
The corresponding set of equilibrium solutions $\xi=\xi_*$ for different values of $\alpha_{\rm sc}$ is listed in Table \ref{tab:1}.


\begin{table}
\centering
\caption{The pair of equilibrium solutions $\xi=\xi_*^\pm$ is shown for different values of $\alpha_{\rm sc}$ greater than the critical value $\alpha_{\rm sc}^{\rm crit}\approx2.4554$ such that $\xi_*^-\simeq\xi_*^+\approx0.5117$. Note  that $\xi\to\xi_*^+$ at late times.
}
\begin{ruledtabular}
\begin{tabular}{lll}
$\alpha_{\rm sc}$ & $\xi_*^{-}$ & $\xi_*^{+}$ \\
\hline
2.45540748& 0.51169966& 0.51169969\\
2.4555& 0.50659526& 0.51680963\\
2.456& 0.49879508& 0.52463974\\
2.46& 0.47591246& 0.54776068\\
2.5& 0.40262914& 0.62332204\\
3& 0.19258836& 0.84988278\\
3.5& 0.12405732& 0.92230203\\
4& 0.08793621& 0.95696028\\
4.5& 0.06596326& 0.97532713\\
5& 0.05145050& 0.98556820\\
5.5& 0.04131315& 0.99145133\\
6& 0.03393271& 0.99489482\\
6.5& 0.02838350& 0.99693497\\
7& 0.02410150& 0.99815336\\
7.5& 0.02072581& 0.99888485\\
8& 0.01801619& 0.99932556\\
9& 0.01398286& 0.99975265\\
10& 0.01117017& 0.99990912\\
15& 0.00477222& 0.99999939\\
\end{tabular}
  \end{ruledtabular}
\label{tab:1}
\end{table}

The late-time evolution of the effective equation of state parameter (\ref{weffdef}) is thus given by
\begin{eqnarray}
w_{\rm eff}&=&w_{\rm(in)}\left(1+\frac{g}{2} \alpha_{\rm sc}^2\xi\right)
+{\rm O}(\xi^2), \qquad \xi\to0, \nonumber\\
w_{\rm eff}&=&w_{\rm(in)}\left(1+\frac{g}{2} \frac{\psi_*^2}{\xi_*}\right)
+{\rm O}(\xi-\xi_*), \qquad \xi\to\xi_{*},
\end{eqnarray}
where $\psi_*=\psi(\xi_*)$, so that $w_{\rm eff}\to w_{\rm(in)}$ for $\xi\to0$, while $w_{\rm eff}\to-1$ for $\xi\to\xi_*$.
Therefore, the SC models evolve towards a universe filled by an ordinary fluid with barotropic equation 
of state $p=w_{\rm(in)}\rho$ in the former case, whereas in the latter case they approach a de Sitter universe with $p=-\rho=-\rho_*$.
The stability properties of the attractor solutions are discussed in Appendix \ref{appstab}.

\subsection{Scalar field description}

Let the SC fluid be dynamically implemented by a homogeneous scalar field $\phi(t)$ 
minimally coupled to gravity, with energy density $\rho$ and pressure $p$ related to the kinetic and potential energy in a standard way, i.e.,
\begin{equation}
\label{rhoandp}
\rho=\frac{\dot\phi^2}{2}+V,\qquad
p=\frac{\dot\phi^2}{2}-V,
\end{equation}
so that
\begin{equation}
\label{Vdef}
V=\frac12(\rho-p),\qquad
\dot\phi^2=\rho+p.
\end{equation}
The field equations (\ref{FRWeqs}) imply that the dynamics of the scalar field is governed by the equation
\begin{equation}
\ddot\phi+3H\dot\phi+V'=0,
\end{equation}
where a prime denotes differentiation with respect to the scalar field $\phi$.

\begin{figure}
\centering
\subfigure[]{\includegraphics[scale=0.33]{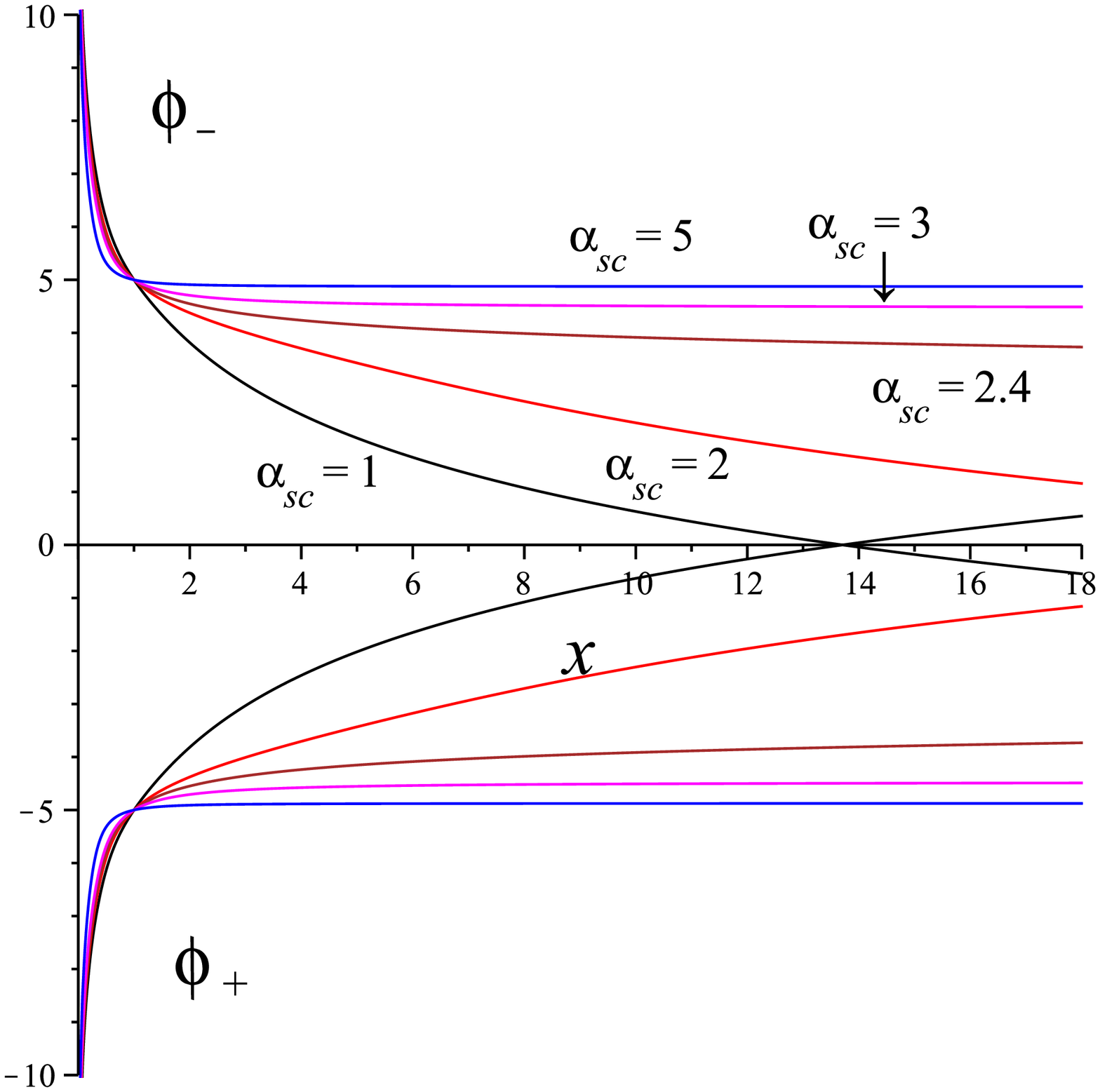}}
\hspace{5mm}
\subfigure[]{\includegraphics[scale=0.35]{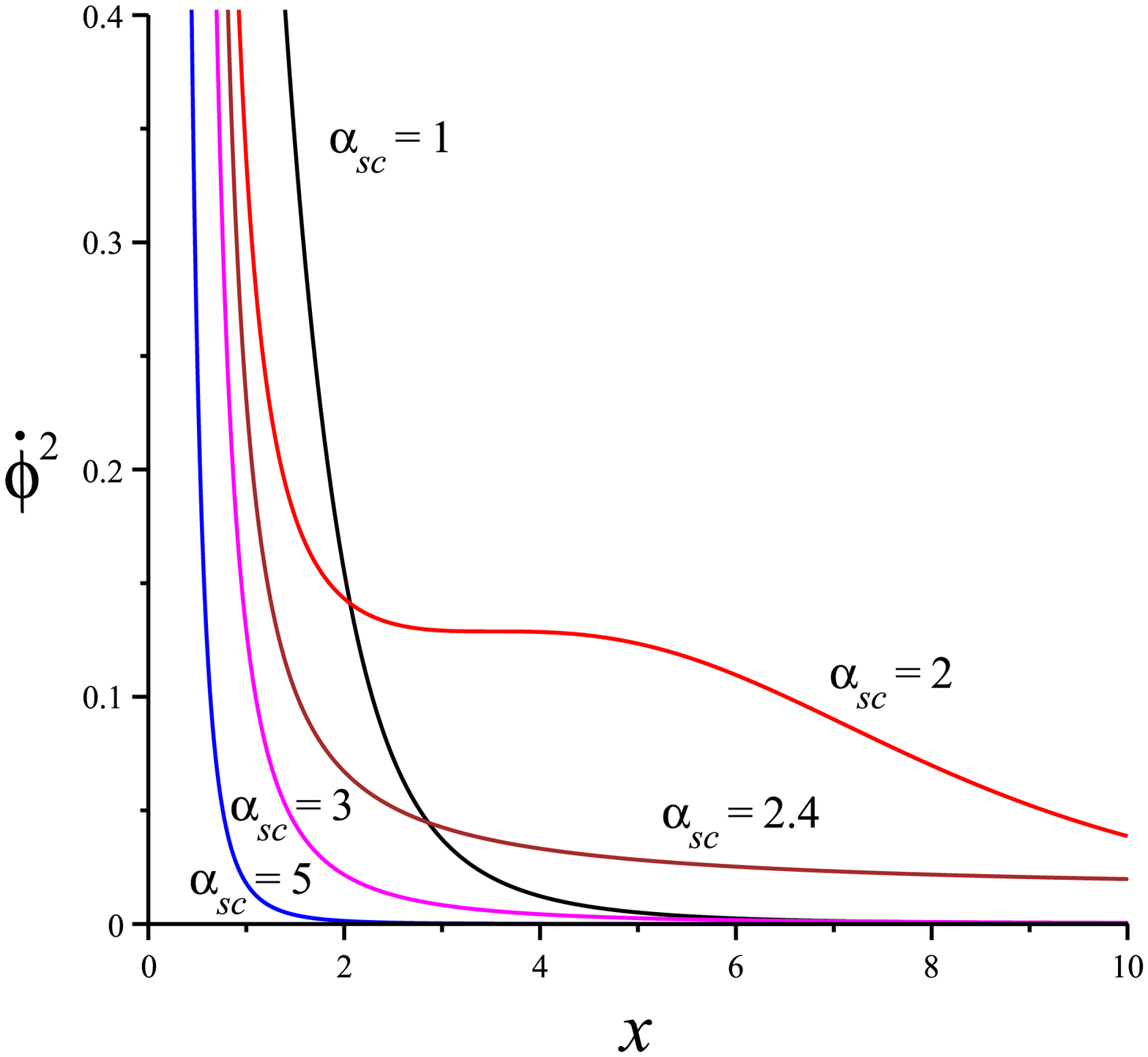}}
\caption{
The behavior of $\phi$ (in units of $\kappa$, panel (a)) and $\dot \phi^2$ (in units of $\rho_{\rm (crit),0}$, panel (b)) as a function of the dimensionless scale factor $x$ is shown for a 
SC model with the same choice of parameters as in Fig. \ref{fig:weff}.
Initial conditions are chosen as in Fig. \ref{fig:weff}, with in addition $\phi_-(\tau_0)=-\phi_+(\tau_0)=5$.
The two branches $\phi_+$ (lower) and $\phi_-$ (upper) asymptotically either diverge crossing each other 
($\phi_\pm\to\pm\infty$) or approach a finite constant value for $\alpha_{\rm sc}\gtrapprox2.455$.
Notice that the labels $\pm$ refer to their asymptotic character in the diverging case.
In particular, $\lim_{x\to\infty}\phi_\pm=\phi^*_\pm\approx\mp4.47$ for $\alpha_{\rm sc}=3$, 
and $\phi^*_\pm\approx\mp4.87$ for $\alpha_{\rm sc}=5$.
The quantity $\dot \phi^2$, instead, vanishes asymptotically for every fixed value of $\alpha_{\rm sc}$.
}
\label{fig:phi_phidot}
\end{figure}

\begin{figure}
\centering
\subfigure[]{\includegraphics[scale=0.35]{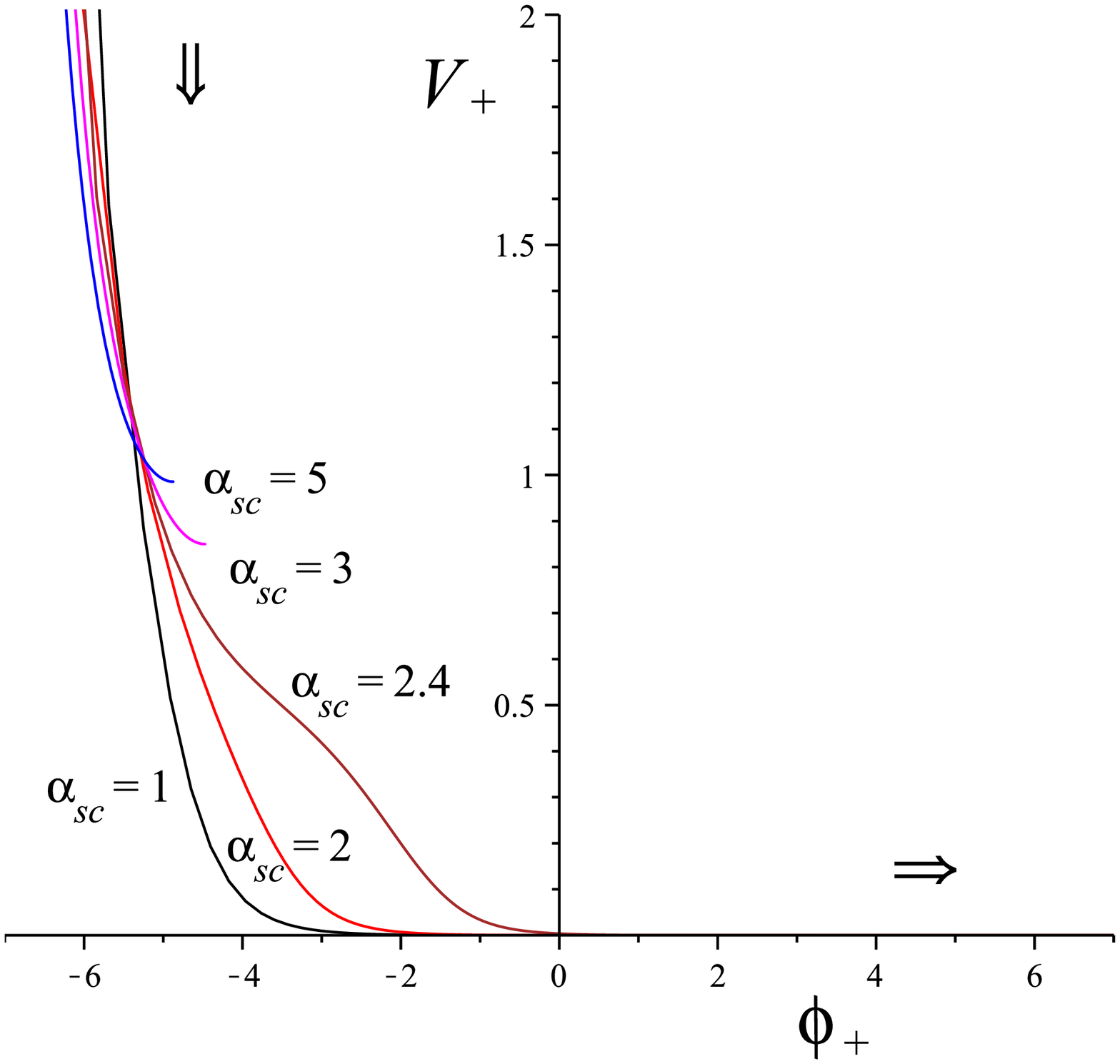}}
\hspace{5mm}
\subfigure[]{\includegraphics[scale=0.35]{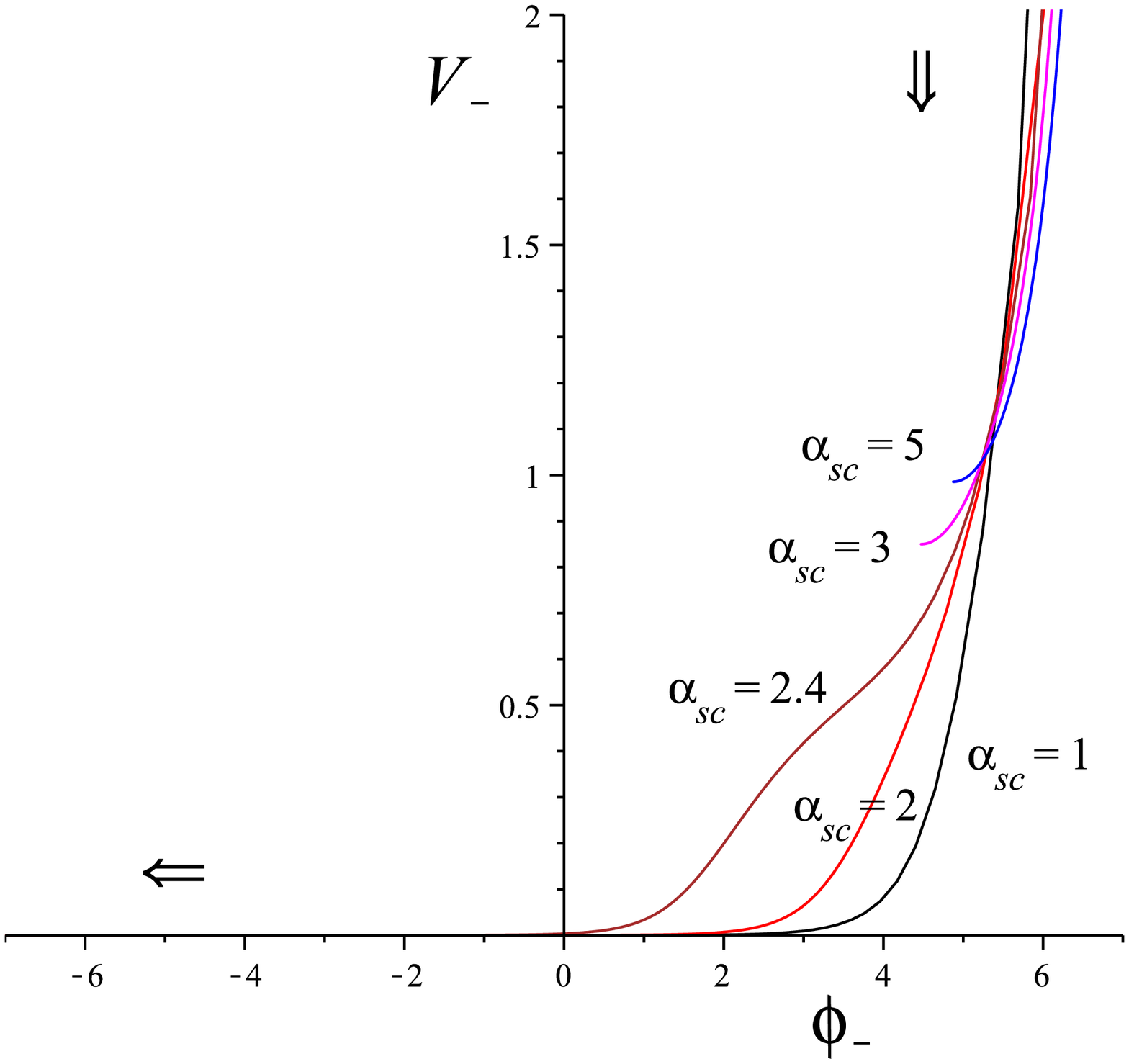}}
\caption{
The behavior of the SC potentials $V_\pm$ (in units of $\rho_{\rm (crit),0}$, panels (a) and (b), respectively) corresponding to the two branches 
$\phi_\pm$ (in units of $\kappa$) of the scalar field is shown for the same choice as well as initial conditions as in Fig. \ref{fig:phi_phidot}.
Arrows indicate the direction of approach to the asymptotic regime. 
}
\label{fig:Vdiphi}
\end{figure}

In terms of the dimensionless variables (\ref{adimvar}) the potential takes the form
\begin{equation}
\label{potdef}
\frac{V}{\rho_{\rm (crit),0}}=\frac12\left(\xi-w_{\rm (in)}{\mathcal P}(\xi)\right),
\end{equation}
whereas the evolution equation for the scalar field reads as
\begin{equation}
\label{eqphi}
\kappa\frac{{\rm d} \phi}{{\rm d}\tau}=\pm\sqrt{3(\xi+w_{\rm (in)}{\mathcal P})}.
\end{equation}
An example of numerical integration of the model equations (\ref{FRWeqs2}) and (\ref{eqphi}) is shown in Figs. \ref{fig:phi_phidot} and \ref{fig:Vdiphi} for the same choice of parameters as well as initial conditions as in Fig. \ref{fig:weff}.

Figure \ref{fig:phi_phidot} shows the behavior of $\phi$ and $\dot \phi^2$ as a function of the dimensionless scale factor 
$x$ for selected values of the parameters. The scalar field has two branches, each of them either extending to 
infinity ($\phi_\pm\to\pm\infty$) or approaching a finite constant value ($\phi_\pm\to\phi_\pm^*$) for large $x$, depending on the chosen value of 
$\alpha_{\rm sc}$ (for fixed $w_{\rm (in)}$ and $g$).
The associated kinetic energy is always positive in the entire domain and vanishes asymptotically.

Figure \ref{fig:Vdiphi} shows, instead, the profile $V=V(\phi)$ of the SC potential.
For every fixed value of $\alpha_{\rm sc}$, the left (right) branch $V_+$ ($V_-$) corresponds to $\phi_+$ ($\phi_-$).
Therefore, the potential either vanishes asymptotically ($V_\pm\to0$ for $\phi_\pm\to\pm\infty$) or 
tends to a constant ($V_\pm\to V_*$ for $\phi_\pm\to\phi_\pm^*$).


\subsection{Asymptotic profile of the SC potential}

In order to find an explicit form of the potential when approaching the asymptotic regime, 
let us solve the SC dynamical equations in this limit. They are given by
\begin{eqnarray}
\label{final_sys}
\frac{{\rm d}\xi}{{\rm d}\alpha}&=&-\lambda^2\xi\left[1+\frac32\frac{w_{\rm (in)}g}{\lambda^2}\frac{\psi^{2}(\xi)}{\xi}\right],
\nonumber\\
\kappa\frac{{\rm d}\phi_\pm}{{\rm d}\alpha}&=&\pm\lambda\sqrt{1+\frac32\frac{w_{\rm (in)}g}{\lambda^2}\frac{\psi^{2}(\xi)}{\xi}},
\end{eqnarray}
where $\alpha=\ln x$ such that ${{\rm d}\alpha}/{{\rm d}\tau}=\sqrt{\xi}$ and $\lambda^2=3(1+w_{\rm (in)})>0$, since   
we are considering the case of a scalar field which behaves as an ordinary field in both low and high density regimes only, so that $w_{\rm (in)}>-1$. The special case $w_{\rm (in)}=-1$ will be treated separately.

According to the analysis done previously, if $\alpha_{\rm sc}<\alpha_{\rm sc}^{\rm crit}$ the energy density vanishes asymptotically ($\xi\to0$), so that $\psi(\xi)=\alpha_{\rm sc}\xi+{\rm O}(\xi^2)$, whereas the scalar field is positively/negatively diverging with $\alpha$.
In fact, in this limit Eq. (\ref{final_sys}) gives the equilibrium solution
\beq
\label{equil1}
\xi=0,\qquad
\kappa(\phi_\pm-\phi_\pm^0)=\pm\lambda(\alpha-\alpha_0),
\eeq
where one can set $\phi_\pm^0=0=\alpha_0$ without any loss of generality.
If $\alpha_{\rm sc}>\alpha_{\rm sc}^{\rm crit}$, instead, both the energy density and the scalar field tend to a constant value ($\xi\to\xi_*$, $\phi\to\phi_\pm^*$), so that $\psi(\xi)=\psi_*+(1-\psi_*)\alpha_{\rm sc}(\xi-\xi_*) +{\rm O}[(\xi-\xi_*)^2]$ and the equilibrium solution is
\beq
\label{equil2}
\xi=\xi_*,\qquad
\phi_\pm=\phi_\pm^*.
\eeq

\subsubsection{Case 1: $\alpha_{\rm sc}<\alpha_{\rm sc}^{\rm crit}$}

Linear perturbations $\xi=\epsilon\xi_{\rm(1)}$ and $\kappa\phi_\pm=\pm\lambda\alpha+\epsilon\kappa\phi_{\rm(1)}$ of Eq. (\ref{final_sys}) around the equilibrium solution (\ref{equil1}) give
\beq
\frac{{\rm d}\xi_{\rm(1)}}{{\rm d}\alpha}=-\lambda^2\xi_{\rm(1)},\qquad
\kappa\frac{{\rm d}\phi_{\rm(1)}}{{\rm d}\alpha}=\mp\lambda\sigma\xi_{\rm(1)},
\eeq
where $\epsilon$ denotes a smallness indicator and 
\beq
\sigma=-\frac34\frac{w_{\rm (in)}g\alpha_{\rm sc}^2}{\lambda^2},
\eeq
implying that 
\beq
\label{solord1}
\xi_{\rm(1)}=\bar\xi\,{\rm e}^{-\lambda^2\alpha},\qquad
\kappa\phi_{\rm(1)}=\pm\frac{\sigma}{\lambda}\xi_{\rm(1)}.
\eeq
Inverting the relation $\phi=\phi(\alpha)$ to the lowest order then gives
\beq
\label{solxiord1}
\xi=\bar\xi\,{\rm e}^{\mp\kappa\lambda\phi_\pm},
\eeq
where $\bar\xi$ is an integration constant and we have set $\epsilon=1$.

The above first order solution does not depend on the SC parameters $g$ and $\alpha_{\rm sc}$.
Therefore, it is worth going to the next order looking for solutions of the form $\xi=\epsilon\xi_{\rm(1)}+\epsilon^2\xi_{\rm(2)}$ and $\kappa\phi_\pm=\pm\lambda\alpha+\epsilon\kappa\phi_{\rm(1)}+\epsilon^2\kappa\phi_{\rm(2)}$.
Equations (\ref{final_sys}) then imply 
\beq
\frac{{\rm d}\xi_{\rm(2)}}{{\rm d}\alpha}=-\lambda^2\left(\xi_{\rm(2)}-2\sigma\xi_{\rm(1)}^2\right),\qquad
\kappa\frac{{\rm d}\phi_{\rm(2)}}{{\rm d}\alpha}=\mp\lambda\sigma\left[\xi_{\rm(2)}-\left(\alpha_{\rm sc}-\frac{\sigma}2\right)\xi_{\rm(1)}^2\right].
\eeq
Using the first order solution (\ref{solord1}) we find
\beq
\xi_{\rm(2)}=-2\sigma\bar\xi^2\,{\rm e}^{-2\lambda^2\alpha},\qquad
\kappa\phi_{\rm(2)}=\pm\frac{2\alpha_{\rm sc}+3\sigma}{2\lambda}\xi_{\rm(2)}.
\eeq
Expressing $\alpha$ in terms of $\phi$ by inverting perturbatively the relation $\phi=\phi(\alpha)$ finally leads to
\beq
\label{solxiord2}
\xi=\bar\xi\,{\rm e}^{\mp\kappa\lambda\phi_\pm}\left(1-\sigma\bar\xi\,{\rm e}^{\mp\kappa\lambda\phi_\pm}\right).
\eeq

Note that the solution (\ref{solxiord2}) can also be obtained by expanding up to the second order in $\xi$ both equations (\ref{final_sys}), which can be combined to yield
\beq
\frac1{\kappa}\frac{{\rm d}\xi}{{\rm d}\phi_\pm}=\mp\lambda\xi(1-\sigma\xi),
\eeq
with solution
\begin{equation}
\xi=\left(\frac1{\bar\xi}{\rm e}^{\pm\lambda\kappa\phi_\pm}+\sigma\right)^{-1}.
\end{equation}
Therefore, the profile of the potential 
\begin{equation}
\frac{V}{\rho_{\rm (crit),0}}
=\frac12\xi(1-w_{\rm (in)})+{\rm O}(\xi^{2})
\end{equation}
is of the form
\begin{equation}
\label{Vcase1}
V(\phi)=\frac{V_0}{{\rm e}^{\lambda\kappa\phi}+B}, \qquad
B=\sigma\bar\xi, \qquad
V_0=\frac12\bar\xi(1-w_{\rm (in)})\rho_{\rm (crit),0},
\end{equation}
where we have dropped the sign indicator, since the $\pm$ signs correlate with those of $\phi_+$/$\phi_-$ which are positively/negatively increasing during the evolution.
Very close to the attractor solution one can neglect the constant term $B$, so that the 
SC potential is well approximated by a decreasing exponential, i.e.,
\begin{equation}
\label{Vcase1exp}
V(\phi)=V_0\, {\rm e}^{-\lambda\kappa\phi},
\end{equation}
which corresponds to the lowest order solution (\ref{solxiord1}).
Therefore, in this regime the SC fluid exhibits the same features of an ordinary fluid with a barotropic equation of state $p=w_{\rm(in)}\rho$ with constant parameter $w_{\rm(in)}$.

\subsubsection{Case 2: $\alpha_{\rm sc}>\alpha_{\rm sc}^{\rm crit}$}

\begin{figure}
\begin{center}
\includegraphics[scale=0.35]{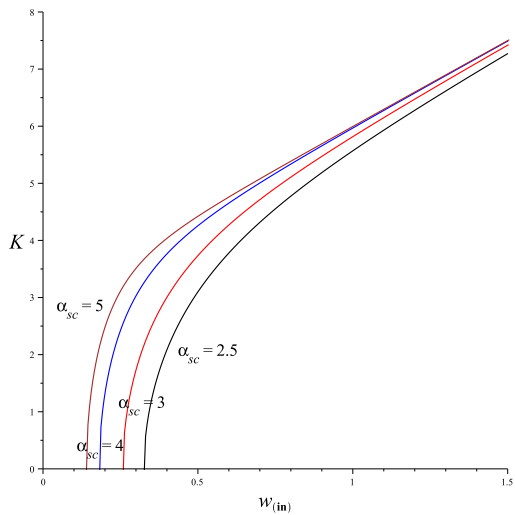}
\end{center}
\caption{
The behavior of the quantity $K$ given by Eq. (\ref{Kdef}) as a function of $w_{\rm(in)}$ is shown for $g=-8$ and different values of $\alpha_{\rm sc}=[2.5,3,4,5]$. The values of $\xi_*$ corresponding to each pair $(\alpha_{\rm sc},w_{\rm(in)})$ are computed by Eq. (\ref{eq_F}).
}
\label{fig:K}
\end{figure}

Linear perturbations $\xi=\xi_*+\epsilon\xi_{\rm(1)}$ and $\phi_\pm=\phi_\pm^*+\epsilon\phi_{\rm(1)}$ of Eq. (\ref{final_sys}) around the equilibrium solution give
\beq
\frac{{\rm d}\xi_{\rm(1)}}{{\rm d}\alpha}=-K\xi_{\rm(1)},\qquad
\kappa\frac{{\rm d}\phi_{\rm(1)}}{{\rm d}\alpha}=0,
\eeq
implying that
\begin{equation}
\xi\simeq\xi_*+\bar\xi {\rm e}^{-K\alpha},\qquad
\phi_\pm\simeq\phi_\pm^{*} ,
\end{equation}
where $\bar\xi$ is an integration constant and
\begin{equation}
\label{Kdef}
K=\lambda^2(1+2\alpha_{\rm sc}\xi_*)+3w_{\rm (in)}g\alpha_{\rm sc}\psi_{*},
\end{equation}
which has to be positive to ensure stability. Its behavior as a function of $w_{\rm(in)}$ is shown in Fig. \ref{fig:K} for $g=-8$ and different values of $\alpha_{\rm sc}$ (for instance, for $\alpha_{\rm sc}=3$ and $w_{\rm(in)}=1/3$ we get $\xi_*\approx0.85$ and $K\approx2.27$).
As a result, the potential does not depend on $\phi$, but is a function of $\alpha$  
\begin{equation}
\frac{V}{\rho_{\rm (crit),0}}
\simeq\xi_*+\bar\xi\left(1+\frac{K}{6}\right){\rm e}^{-K\alpha},
\end{equation}
and the corresponding profile is
\begin{equation}
\label{Vcase2}
V(\alpha)=V_*\left(1+q{\rm e}^{-K\alpha}\right), \qquad
q=\frac{\bar\xi}{\xi_*}\left(1+\frac{K}{6}\right), \qquad
V_*=\xi_{*} \rho_{\rm (crit),0}.
\end{equation}
For very large values of the scale factor, one can neglect the exponential term, so that the 
SC potential is well approximated by a constant, i.e., $V(\alpha)=V_{*}$.
Therefore, in this regime the SC fluid acts as a cosmological constant.


\subsubsection{Case 3: $w_{\rm (in)}=-1$}

For $w_{\rm (in)}=-1$ the dynamical equations (\ref{final_sys}) reduce to
\beq
\label{syswineqm1}
\frac{{\rm d}\xi}{{\rm d}\alpha}=-\frac32|g|\psi^{2}(\xi),\qquad
\kappa\frac{{\rm d}\phi_\pm}{{\rm d}\alpha}=\pm\sqrt{\frac32|g|}\,\frac{\psi(\xi)}{\sqrt{\xi}}.
\eeq
The equation for $\xi$ can be solved exactly (see also Ref. \cite{scinfla})
\beq
\xi=\frac1{\alpha_{\rm sc}}\ln\left(1+\frac1y\right), \qquad
y=W\left(y_0\,{\rm e}^{\tilde\sigma\alpha+y_0}\right),
\eeq
where $\tilde\sigma=\frac32|g|\,\alpha_{\rm sc}$ and $W(z)$ denotes the principal branch of the Lambert $W$ function such that $W(z){\rm e}^{W(z)}=z$. Its asymptotic behavior for $z\to\infty$ is $W(z)\sim\ln z$, implying that for large values of the scale factor the solution behaves as $y\sim\tilde\sigma\alpha$ and $\alpha_{\rm sc}\xi\sim1/(\tilde\sigma\alpha)$. Substituting then into the equation for $\phi=\phi_\pm$ gives $\kappa\phi_\pm\sim\pm2\sqrt{\alpha}$, so that $\xi\sim4/(\tilde\sigma\alpha_{\rm sc}\kappa^2\phi^2)$.
Equivalently, the solution for $\xi$ as a function of $\phi$ can be obtained by re-expressing Eq. (\ref{syswineqm1}) as
\beq
\frac1{\kappa}\frac{{\rm d}\xi}{{\rm d}\phi_\pm}=\mp\sqrt{\frac32|g|\xi}\,\psi(\xi).
\eeq
Close to the attractor solution $\xi\to0$ the above equation becomes
\beq
\frac1{\kappa}\frac{{\rm d}\xi}{{\rm d}\phi_\pm}\simeq\mp\sqrt{\tilde\sigma\alpha_{\rm sc}}\xi^{3/2},
\eeq
with solution
\beq
\xi\simeq\left[\frac1{\sqrt{\bar\xi}}\pm\frac12\sqrt{\tilde\sigma\alpha_{\rm sc}}\kappa\phi_\pm\right]^{-2},
\eeq
where $\bar\xi$ is an integration constant such that $\xi=\bar\xi$ for $\phi=0$.
Therefore, the associated potential 
\beq
\frac{V}{\rho_{\rm (crit),0}}=\xi+{\rm O}(\xi^{2})
\eeq
is an inverse square function of $\phi$ at late times (where $\phi_\pm\to\pm\infty$), i.e., 
\beq
\label{Vcase3}
V(\phi)\simeq\frac{V_0}{(\kappa\phi)^{2}},\qquad
V_0=\frac{4}{\tilde\sigma\alpha_{\rm sc}}\rho_{\rm (crit),0}.
\eeq

\section{Quantum features}

Let us study now the quantum features of the late-time evolution of SC cosmological models described in Sec. II. 
We will follow the approach of quantum geometrodynamics based on the WDW equation, which reads 
${\mathcal L}_{\rm (wdw)}\Psi(\alpha, \phi)=0$, where
\begin{equation}
\label{wdweq}
{\mathcal L}_{\rm (wdw)}=
\frac{\hbar^2}{2}\left[\frac{\kappa^2}{6}\partial_{\alpha\alpha}-\partial_{\phi\phi}\right]
+a_0^6 {\rm e}^{6\alpha} V(\phi),
\end{equation}
with $\kappa^2 \equiv 8\pi G$ and $\alpha \equiv \ln(a/a_0)$, as specified before.
Hereafter, we shall set $a_0=1$ and $\kappa^2=6$, for simplicity.
As a second order differential equation (of the hyperbolic type in minisuperspace applications), 
the WDW equation requires two data at the Cauchy surface, i.e., the value of the function and its \lq\lq time'' derivative.
A way to select physically meaningful solutions for the wave function $\Psi(\alpha, \phi)$ matching with quantum initial value data has been recently proposed in Ref. \cite{kamen89}. 
At the semiclassical level, these solutions are 
represented by oscillating waves in classically allowed domains of superspace, 
and exponential fall off in classically forbidden regions.

The WDW equation can be explicitly solved only for particular potentials or under some approximation.
We will adopt the BO approximation, briefly reviewed in Appendix \ref{appBOapprox} (see, e.g., Ref. \cite{kiefer88}), so that
\begin{equation}
\Psi(\alpha, \phi)=\sum_k \phi_k(\alpha,\phi)C_k(\alpha),
\label{PsiBO}
\end{equation}
where $\phi_{k}$ and $C_{k}$ are the slow and fast degrees of freedom, respectively, satisfying Eq. (\ref{(3.9)}).
Since we are interested in the asymptotic behavior towards future attractor solutions, we will limit our 
investigations to approximate potentials in the neighborhood of such attractors.
Therefore, we will consider below the two asymptotic SC potentials (\ref{Vcase1}), (\ref{Vcase2}) and (\ref{Vcase3}), corresponding to Case 1, Case 2  and Case 3, respectively, discussed in Section II B.
Very close to the attractor solutions the above potentials are well approximated by an exponential potential and a constant potential, respectively, which allow to find exact solutions to the WDW equation, as recalled in Appendix \ref{appWDWexppot}.

\subsection{Case 1}

Let us consider first the potential (\ref{Vcase1}).
In the BO approximation, the equation for $\phi_k$ reads
\begin{equation}
\label{eqphik}
\hbar^2\phi_k'' +2 \left[E_{k}(\alpha)-\frac{V_0{\rm e}^{6\alpha}}{{\rm e}^{\lambda\kappa\phi}+B}\right]\phi_k=0,
\end{equation}
which can be cast in the form of a hypergeometric equation
\begin{equation}
\label{hypeq}
\left[z(1-z)\frac{{\rm d}^2}{{\rm d}z^2} +[c-(a+b+1)z]\frac{{\rm d}}{{\rm d}z}-ab
\right] f_{k}=0,
\end{equation}
through the transformation
\begin{equation}
\phi_k = z^{-e_{k}(\alpha)/2}(z-1)f_k,\qquad
z = -B\,{\rm e}^{-\lambda\kappa\phi},
\end{equation}
with
\begin{equation}
a = 1-\frac12[b_{k}(\alpha)+e_{k}(\alpha)],\qquad
b = 1+\frac12[b_{k}(\alpha)-e_{k}(\alpha)],\qquad
c = 1-e_{k}(\alpha)=a+b-1,
\end{equation}
and
\begin{equation}
e_{k}(\alpha) = \frac{2\sqrt{-2E_{k}(\alpha)}}{\lambda\kappa\hbar},\qquad
b_{k}(\alpha)^2 =e_{k}(\alpha)^2
+\frac{8V_0{\rm e}^{6\alpha}}{\lambda^2\kappa^2\hbar^2 B}.
\end{equation}
Equation (\ref{hypeq}) has general solution
\begin{equation}
f_k=c_1(\alpha)F(a,b,c;z)+c_2(\alpha)z^{1-c}F(1+a-c,1+b-c,2-c;z).
\end{equation}
The latter can be considered as a special case of the confluent Heun function, already shown to play a role in this context in Ref. \cite{kiefer3}, i.e.,
\begin{eqnarray}
f_k&=&\tilde c_1(\alpha)(z-1)^{-a}{\rm HeunC}\left(0,b_{k}(\alpha),-e_{k}(\alpha),0,\frac14(b_{k}(\alpha)^2+e_{k}(\alpha)^2);\frac1{1-z}\right)\nonumber\\
&&
+\tilde c_2(\alpha)(z-1)^{-b}{\rm HeunC}\left(0,-b_{k}(\alpha),-e_{k}(\alpha),0,\frac14(b_{k}(\alpha)^2+e_{k}(\alpha)^2);\frac1{1-z}\right).
\end{eqnarray}

We are interested in normalizable solutions, so as to obtain a discrete spectrum.
Therefore, we will consider only those solutions such that the hypergeometric functions reduce to orthogonal 
polynomials, including Jacobi polynomials $P_k^{(\gamma,\delta)}$ and their special cases 
(Legendre or Chebyshev or Gegenbauer polynomials), by using the relation
\begin{equation}
\label{jacobi}
F(-k,\gamma+\delta+1+k,\gamma+1;z)=\frac{k!}{(\gamma+1)_k}P_k^{(\gamma,\delta)}(1-2z),
\end{equation}
where $(m)_{n} \equiv (m+n-1)!/(m-1)!$ is a Pochhammer symbol. They satisfy the orthogonality condition
\begin{equation}
\int_{-1}^1(1-x)^\gamma(1+x)^\delta P_j^{(\gamma,\delta)}(x)P_k^{(\gamma,\delta)}(x)dx
=\frac{2^{\gamma+\delta+1}}{2k+\gamma+\delta+1}\frac{\Gamma(k+\gamma+1)
\Gamma(k+\delta+1)}{\Gamma(k+\gamma+\delta+1)k!}\delta_{jk},
\end{equation}
with $\gamma,\delta,\gamma+\delta>-1$, and are normalized in such a way that
\begin{equation}
P_k^{(\gamma,\delta)}(1)={{k+\gamma}\choose{k}}.
\end{equation}
Therefore, in the limit of large $\phi$ (implying $z\to0$) the functions $f_k$ chosen as the Jacobi polynomials 
of Eq. (\ref{jacobi}) take a constant value.  Equation (\ref{jacobi}) then implies
\begin{equation}
\gamma=-e_{k}(\alpha),\qquad
\delta=1,\qquad
2(k+1)=e_{k}(\alpha)-b_{k}(\alpha),\qquad
e_{k}(\alpha)=k+1-\frac{2V_0{\rm e}^{6\alpha}}{(k+1)\lambda^2\kappa^2\hbar^2 B},
\end{equation}
and the equation for the functions $C_k(\alpha)$ becomes
\begin{equation}
\label{eqCk}
\left[{{\rm d}^{2}\over {\rm d}\alpha^{2}}-\frac14 \lambda^2\kappa^2 e_k(\alpha)^2 \right] C_k=0.
\end{equation}
This equation can be solved in terms of the Whittaker functions of the first and second 
kind $M_{\mu,\nu}(x)$ and $W_{\mu,\nu}(x)$ as
\begin{equation}
C_k={\rm e}^{-3\alpha}\left[c_1M_{\mu,\nu}(x_k)+c_2W_{\mu,\nu}(x_k)\right],
\end{equation}
with
\begin{equation}
\mu = \frac1{12}(k+1)\lambda\kappa=\nu,\qquad
x_k = \frac{V_0{\rm e}^{6\alpha}}{3(k+1)\lambda\kappa\hbar^2 B}.
\end{equation}
Regularity for large $\alpha$ requires $c_1=0$, and one can set $c_2=1$ without any loss of generality. 
The complete solution is thus given by
\begin{equation}
\Psi(\alpha, \phi)=\sum_k \frac{N_kk!}{(\gamma+1)_k}{\rm e}^{-3\alpha}
W_{\mu,\nu}(x_k)z^{\gamma/2}(z-1)P_k^{(\gamma,\delta)}(1-2z),
\end{equation}
where $N_k$ is a normalization factor.

In order to calculate the wave packets and study their behavior close to the classical solution we need the 
Jeffreys-Wentzel-Kramers-Brillouin (JWKB) solution of Eqs. (\ref{eqphik}) and (\ref{eqCk}), which is given by Eqs. (\ref{JWKBsol})--(\ref{PsiWKB}) with phase factors
\begin{equation}
S^\phi_k(\alpha,\phi)=\hbar\int_{y_0}^y {\rm d}y' \left[-\frac{e_k(\alpha)^2}{4y'^2}
-\frac{(1+k)(1+k-e_k(\alpha))}{y'(1+y')}\right]^{1/2},\qquad
y \equiv \frac{B}{\lambda\kappa}{\rm e}^{-\lambda\kappa\phi},
\end{equation}
and
\begin{equation}
S^\alpha_k(\alpha)=\frac{{\rm i}}{2}(k+1)\lambda\kappa\hbar(\alpha-\alpha_0)
-\frac{{\rm i}}{6}\frac{V_0}{(k+1)\lambda\kappa\hbar B}({\rm e}^{6\alpha}-{\rm e}^{6\alpha_0}).
\end{equation}
Initial conditions are such that $\phi_k(\alpha_0,\phi_0)=1$ and $C_k(\alpha_0)=1$.
Recalling that in the classical regime there exist two different branches $\phi=\phi_\pm$, 
$\Psi(\alpha_0, \phi)$ will be a superposition of two Gaussians 
$\Psi_\pm(\alpha_0, \phi)$, centered at $\phi_\pm^0$, i.e.,
\begin{equation}
\Psi(\alpha_0, \phi)=c_+\Psi_+(\alpha_0, \phi)+c_-\Psi_-(\alpha_0, \phi)\,,\qquad
\Psi_\pm(\alpha_0, \phi)={\rm e}^{-(\phi-\phi_\pm^0)^2/2\sigma^2(\alpha_0)}.
\end{equation}
The corresponding amplitudes $A_k^\pm$ can be obtained by decomposing each Gaussian into the basis of 
eigenfunctions $\phi_k(\alpha_0,\phi)$ obtained above in terms of Jacobi polynomials, i.e.,
\begin{equation}
\Psi_\pm(\alpha_0, \phi)=\sum_kA_k^\pm\phi_k(\alpha_0,\phi)\,,\qquad
A_k^\pm=\int {\rm d} \phi \; {\rm e}^{-(\phi-\phi_\pm^0)^2/2\sigma^2(\alpha_0)}\phi_k(\alpha_0,\phi),
\end{equation}
so that the solution for the wave packet finally writes as
\begin{equation}
\Psi(\alpha, \phi)=\sum_k[c_+A_k^++c_-A_k^-]{\rm e}^{({\rm i}/\hbar)S_k(\alpha,\phi)}.
\end{equation}

\begin{figure}
\begin{center}
\includegraphics[scale=0.45]{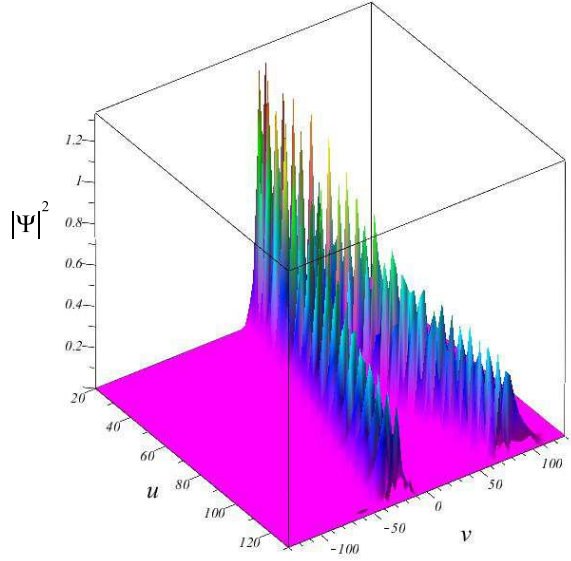}
\end{center}
\caption{The behavior of the (not normalized) squared modulus of the wave packet is shown as a function of the coordinates $(u,v)$ 
for an exponential potential solution of the WDW equation with $\lambda=2$ and 
$V_0/\rho_{\rm (crit),0}\approx1.2\times10^{-4}$ corresponding to the set of 
SC parameters $w_{\rm(in)}=1/3$, $g=-8$ and $\alpha_{\rm sc}=2$ (so that $\bar\xi=\xi(\phi=0)\approx3.5\times10^{-4}$). 
Since the classical solution has two branches $\phi_\pm$ (see Fig. \ref{fig:phi_phidot}), the wave packet is the superposition of two solutions 
of the same form (\ref{Psiewbexp}), but with different values of $\bar k=\pm2/\sqrt{3}$.
The remaining parameters are chosen as $c_1=1$, $\hbar=1$ and $\sigma=0.1$ for both Gaussians.
The width of the single Gaussian profile increases quadratically in $v$ \cite{kiefer1} according to the relation $1+\sigma^4\hbar^2(S_0'')^2=1+27\sigma^4\hbar^2v^2$.
The wave function is peaked around the two branches of the classical trajectory, which diverge for increasing values of $u$ and $v$.
Therefore, decoherence effects may be relevant in the region wherein they cross each other, but the approximation adopted to construct the wave packet is no longer valid there. 
}
\label{fig:Psi}
\end{figure}

Very close to the attractor solution the potential (\ref{Vcase1}) is well approximated by the exponential 
potential (\ref{Vcase1exp}), i.e., $V(\phi)=V_0 {\rm e}^{-\lambda\kappa\phi}$.
In this case, the solution in the BO approximation is of the same form as Eq. (\ref{BOsolexp}).
Furthermore, since the classical solution has two branches $\phi_\pm$ (see Fig. \ref{fig:phi_phidot}), the wave packet is actually the superposition of two solutions 
of the same form (\ref{Psiewbexp}), but with different values of $\bar k$.
Therefore, the squared modulus $|\Psi|^2$ will be proportional to the sum of the contributions of the separate packets plus an interference term, i.e.,  $|\Psi|^2\propto|\Psi_1|^2+|\Psi_2|^2+2{\rm Re}(\Psi_1^*\Psi_2)$. The latter may be responsible for decoherence effects \cite{Ha89,kiefer92}. 
The spreading of the wave packet near the classical trajectory is shown in Fig. \ref{fig:Psi}, where the 
behavior of $|\Psi|^2$ as a function of the coordinates $(u,v)$ is plotted for an exponential 
potential solution of the WDW equation with $\lambda=2$ and 
$V_0/\rho_{\rm (crit),0}\approx1.2\times10^{-4}$ corresponding to the set of 
SC parameters $w_{\rm in}=1/3$, $g=-8$ and $\alpha_{\rm sc}=2$.
Since the exponential potential approximation adopted here holds for very large values of $\phi$ (and hence for large values of $u$ and $v$), the two branches of the wave function are well separated and one cannot envisage decoherence effects between them in this regime.


\subsection{Case 2}

Let us consider now the potential (\ref{Vcase2}).
In the BO approximation, the equation for $\phi_k$ reads
\begin{equation}
\hbar^2\phi_k'' +2 \left[E_{k}(\alpha)-V_*{\rm e}^{6\alpha}\left(1+q{\rm e}^{-K\alpha}\right)\right]\phi_k=0.
\end{equation}
Looking for solutions of the form $\phi_k(\alpha, \phi)={\rm e}^{k\phi}f(\alpha)$, we get
\begin{equation}
E_{k}(\alpha)=-\frac{1}{2}\hbar^2k^2+V_*{\rm e}^{6\alpha}\left(1+q{\rm e}^{-K\alpha}\right).
\end{equation}
The equation for the functions $C_k(\alpha)$ then becomes
\begin{equation}
\label{eqCkconst}
\left\{\frac{{\rm d}^2}{{\rm d}\alpha^2}-\left[k^2-\frac{2}{\hbar^2}V_*{\rm e}^{6\alpha}\left(1+q{\rm e}^{-K\alpha}\right)\right]
\right\}C_{k}=0,
\end{equation}
which does not admit explicit solutions.

However, for very large values of the scale factor the potential (\ref{Vcase2}) is well approximated by the 
constant potential $V(\alpha)=V_*$. Therefore, neglecting the term ${\rm e}^{-K\alpha}$ in Eq. (\ref{eqCkconst}) leads to
\begin{equation}
C_k=c_1J_{k \over 3}\left(\frac{\sqrt{2V_*}}{3\hbar}{\rm e}^{3\alpha}\right)
+c_2Y_{k \over 3}\left(\frac{\sqrt{2V_*}}{3\hbar}{\rm e}^{3\alpha}\right),
\end{equation}
where $J_{\nu}(x)$ and $Y_{\nu}(x)$ are the Bessel functions of first and second kind, respectively.
Regularity on the whole axis requires $c_2=0$, so that the general solution is
\begin{equation}
\Psi(\alpha, \phi)=\sum_kN_k{\rm e}^{k\phi}J_{k \over 3}\left(\frac{\sqrt{2V_*}}{3\hbar}{\rm e}^{3\alpha}\right),
\end{equation}
where $N_k$ is a normalization factor.

\subsection{Case 3}

Finally, let us consider the case $w_{\rm (in)}=-1$, with associated potential (\ref{Vcase3}).
In the BO approximation, the equation for $\phi_k$ reads
\begin{equation}
\hbar^2\phi_k'' +2 \left[E_{k}(\alpha)-\frac{V_0{\rm e}^{6\alpha}}{\kappa^2\phi^2}\right]\phi_k=0,
\end{equation}
with general solutions
\beq
\phi_k=\sqrt{|\phi|}\left[c_1(\alpha)J_{\nu}(k|\phi|)+c_2(\alpha)Y_{\nu}(k|\phi|)\right],
\eeq
where
\beq
\nu=\frac12\sqrt{1+\frac{8V_0}{\hbar^2\kappa^2}{\rm e}^{6\alpha}},\qquad
k=\frac{\sqrt{2E_{k}(\alpha)}}{\hbar},
\eeq
so that the index of the Bessel functions depends on $\alpha$.
The quantity $k$ can be real or imaginary depending on whether the energy eigenvalue is positive or negative.
Consider, for instance, the case $k^2>0$ (a more detailed account of inverse square potential in quantum cosmology can be found in Ref. \cite{kiefer4}). 
Regularity on the whole axis requires $c_2=0$. Furthermore, the vanishing of $\phi_k$ for $|\phi|\to0$ and its boundedness for $|\phi|\to\infty$ imply that the $\phi_k$ are orthogonal functions (with $k$ not necessarily an integer). 
The equation for the functions $C_k(\alpha)$ then becomes
\begin{equation}
\left(\frac{{\rm d}^2}{{\rm d}\alpha^2}+k^2\right)C_{k}=0,
\end{equation}
with oscillating solution
\beq
C_{k}=d_1{\rm e}^{ik\alpha}+d_2{\rm e}^{-ik\alpha}.
\eeq
Thus the full solution can be written as
\begin{equation}
\Psi(\alpha, \phi)=\sum_kN_k\sin(k\alpha)\sqrt{|\phi|}J_{\nu}(k|\phi|),
\end{equation}
where $N_k$ is a normalization factor.

\section{Concluding remarks}

We have investigated the properties of cosmological models with fluids obeying a Shan-Chen-like equation of state in the late-time stage of the evolution.
The fluid evolution equations are coupled with Einstein's equations in a spatially flat FRW background. 
The equation of state for the fluid is of the type $p=w_{\rm eff}(\rho)\,\rho$, with $w_{\rm eff}(\rho)$ interpolating between 
constant asymptotic values at large and low energy densities.  
Such an equation of state is widely used in the context of molecular physics as representing systems undergoing phase transitions, and has been recently applied to cosmology to describe a possible scenario for the development of dark energy at the present epoch of our Universe 
as a transition between an ordinary matter state and an \lq\lq exotic'' one. 

Our findings can be summarized as follows.
We have found that there exist two attractor (equilibrium) solutions associated with the SC fluid dynamics at large 
values of the scale factor, depending on the parameters of the model.
The system then evolves towards either a universe filled by an ordinary fluid with barotropic equation of state 
$p=w_{\rm(in)}\rho$ and constant parameter $w_{\rm(in)}$, or approaches a de Sitter universe with $p=-\rho=-\rho_*$.
No future cosmological singularity arises during the evolution.
We have also studied the associated quantum effects in the asymptotic regime in the framework of quantum geometrodynamics 
as described by the WDW equation. For this purpose,
we have adopted here the equivalent scalar field representation of the fluid minimally coupled to gravity, leading 
to an interaction potential, whose profile has been determined by the classical asymptotic SC dynamics. 
We have then solved analytically the associated WDW equation in the BO approximation scheme and 
discussed its main features. 
We have explicitly constructed wave packets and analyzed their spreading around the corresponding classical solutions. 
At the classical level, the scalar field as a function of the scale factor has two branches extending to infinity, which asymptotically either diverge crossing each other or approach a finite constant value.
In the former case the potential is well approximated by a decreasing exponential, whereas in the latter by a constant for very large values of the scale factor. 
Therefore, the wave packet is actually the superposition of two solutions of the WDW equation corresponding to two different sets of initial conditions.
For instance, within the exponential potential approximation to the classical SC dynamics, we have found that the wave function is peaked and spreads around the two branches of the classical trajectory, which are but well separated. As a result, the interference effects are negligible in this regime.

\appendix

\section{Attractor solutions: stability analysis}
\label{appstab}

In order to study the stability properties of the attractor solutions we follow the approach of Ref. \cite{odintsov} (see also Ref. \cite{copeland}), where the background field equations for a two-component cosmological fluid (pressureless matter with density $\rho_{m}=\rho_{m,0}x^{-3}$ plus a dark energy fluid with equation of state $p=p(\rho)$) are written in the form of a plane-autonomous dynamical system as  
\begin{eqnarray}
\label{dyn_sys}
\frac{{\rm d}X}{{\rm d}\alpha}&=&3X(X-Y-c_s^2),
\nonumber\\
\frac{{\rm d}Y}{{\rm d}\alpha}&=&-3X(1-c_s^2)+3Y(1+X-Y),
\end{eqnarray}
where $\alpha\equiv\ln x$, $c_s^2\equiv{\rm d}p/{\rm d}\rho$ is the SC sound speed squared given by 
\beq
c_s^2=w_{\rm(in)}\left[1+g\alpha_{\rm sc}\left(1-{\rm e}^{-\alpha_{\rm sc} \xi }\right){\rm e}^{-\alpha_{\rm sc} \xi }\right],
\eeq
in terms of the dimensionless variables
\beq
X=\frac{\kappa^2}{6H^2}(\rho+p),\qquad
Y=\frac{\kappa^2}{6H^2}(\rho-p),
\eeq
satisfying the Friedmann constraint $\kappa^2\rho_{m}/(3H^2)=1-X-Y$.
We are interested in the late-time regime $x\to\infty$, where the SC fluid dominates over other components ($\rho_{m}\to0$) and the corresponding sound speed approaches a constant value, i.e., $c_s^2\to\bar c_s^2$, with
\begin{eqnarray}
\bar c_s^2&=&w_{\rm(in)}, \qquad \xi\to0, \nonumber\\
\bar c_s^2&=&w_{\rm(in)}\left[1+g\alpha_{\rm sc}\psi_*(1-\psi_*)\right], \qquad \xi\to\xi_{*}.
\end{eqnarray}
The system (\ref{dyn_sys}) thus has fixed points (a) $[X_0,Y_0]=[0,1]$ and (b) $[X_0,Y_0]=[(1+\bar c_s^2)/2,(1-\bar c_s^2)/2]$.

Linear perturbations $X=X_0+\epsilon X_{\rm(1)}$ and $Y=Y_0+\epsilon Y_{\rm(1)}$ around the fixed point give
\begin{eqnarray}
\frac{{\rm d}X_{\rm(1)}}{{\rm d}\alpha}&=&3(2X_0-Y_0-\bar c_s^2)\,X_{\rm(1)}-3X_0\,Y_{\rm(1)},
\nonumber\\
\frac{{\rm d}Y_{\rm(1)}}{{\rm d}\alpha}&=&3(Y_0+\bar c_s^2-1)\,X_{\rm(1)}+3(1+X_0-2Y_0)\,Y_{\rm(1)},
\end{eqnarray}
$\epsilon$ denoting a smallness indicator.
The stability of critical points can be studied by evaluating the associated eigenvalues $m_{1,2}$, which are given by (a) $m_{1,2}=[-3,-3(1+\bar c_s^2)]$ and (b) $m_{1,2}=[3(1+\bar c_s^2),3\bar c_s^2]$.
Therefore, if the evolution is such that $\xi\to0$ asymptotically, then $m_1=-3$ and $m_2=-3(1+w_{\rm(in)})<0$, implying that the solution is a stable node. 
In contrast, if $\xi\to\xi_*$ asymptotically, then the solution is stable for $\bar c_s^2<-1$ only.

\section{Born-Oppenheimer approximation to the WDW equation}
\label{appBOapprox}

Approximate solutions to the WDW equation can be obtained by carrying out a BO approximation \cite{kiefer88}.
Looking for solutions of the form 
\begin{equation}
\Psi(\alpha, \phi)=\sum_k \phi_k(\alpha,\phi)C_k(\alpha),
\label{(3.4)}
\end{equation}
where $\phi_{k}$ and $C_{k}$ are the slow and fast degrees of freedom, respectively (see below),
the WDW equation becomes
\begin{eqnarray}
\sum_k \left \{ \frac{\hbar^2}2 \left[\frac{\kappa^2}{6}(\ddot \phi_k C_k+2\dot \phi_k \dot C_k +\phi_k\ddot C_k) 
-C_k \phi_k''\right] +{\rm e}^{6\alpha} V(\phi) \phi_k C_k \right \}=0,
\label{(3.5)}
\end{eqnarray}
where a dot now denotes the derivative with respect to $\alpha$ and a prime the derivative with respect to $\phi$.
The BO approximation requires that $\phi_k$ satisfies the equation
\begin{equation}
-\frac{\hbar^2}2 \phi_k'' + \left[{\rm e}^{6\alpha} V(\phi)  -E_{k}(\alpha)\right]\phi_k=0,
\label{(3.6)}
\end{equation}
where the quantities $E_{k}(\alpha)$ play the role of eigenvalues of the differential operator 
$-{{\hbar^2} \over 2}{\partial^{2}\over \partial \phi^{2}}+{\rm e}^{6 \alpha}V(\phi)$, leading to
\begin{eqnarray}
\sum_k \left[\frac{\hbar^2}2\frac{\kappa^2}{6}(\ddot \phi_k C_k+2\dot \phi_k \dot C_k +\phi_k\ddot C_k)+E_k(\alpha)\phi_k C_k \right]=0.
\label{(3.7)}
\end{eqnarray}
Assuming then that the functions $C_k$ vary much more rapidly with $\alpha$ than $\phi_k$, one can neglect 
the terms $\ddot \phi_k C_k$ and $\dot \phi_k \dot C_k$, so that the previous equation reduces to
\begin{eqnarray}
\sum_k \left[\frac{\hbar^2}2\frac{\kappa^2}{6}\ddot C_k+E_k(\alpha) C_k \right]\phi_k=0.
\label{(3.8)}
\end{eqnarray}
Therefore, the main equations become 
\begin{eqnarray}
0&=& \hbar^2\phi_k'' +2 \left[E_{k}(\alpha)-{\rm e}^{6\alpha} V(\phi)\right]\phi_k,
\nonumber\\
0&=&  \hbar^2\ddot C_k+2E_k(\alpha) C_k,
\label{(3.9)}
\end{eqnarray}
with a Schr\"{o}dinger-like structure, where we have set $\kappa^2=6$, for simplicity. 
The BO approximation is best fulfilled close to the region of the classical solution.

\subsection{Construction of wave packets}

In order to study semiclassical as well as quantum regimes of the model it is useful to construct wave packets around the classical trajectories \cite{vilenkin,kiefer88}.
One then needs the JWKB solution of Eq. (\ref{(3.9)}), i.e., 
\begin{equation}
\label{JWKBsol}
\phi_k={\rm e}^{({\rm i}/\hbar)S^\phi_k(\alpha,\phi)},\qquad
C_k={\rm e}^{({\rm i}/\hbar)S^\alpha_k(\alpha)},
\end{equation}
where both phase factors $S^\phi_k$ and $S^\alpha_k$ are taken to obey the Hamilton-Jacobi equation to zeroth order in $\hbar$.
The wave packets thus have the general form 
\begin{equation}
\label{PsiWKB}
\Psi(\alpha, \phi)=\sum_kA_k{\rm e}^{({\rm i}/\hbar)S_k(\alpha,\phi)},\qquad
S_k(\alpha,\phi)=S^\phi_k(\alpha,\phi)+S^\alpha_k(\alpha),
\end{equation}
where the amplitude $A_k$ depends on the initial conditions suitably chosen on the hypersurface 
$\alpha=\alpha_0$ corresponding to the classical solution, with $\Psi(\alpha_0, \phi)$ having a Gaussian 
profile with width $\sigma(\alpha_0)=\sigma_0$ centered at $\phi(\alpha_0)=\phi_0$, i.e.,
\begin{equation}
\Psi(\alpha_0, \phi)={\rm e}^{-(\phi-\phi_0)^2/2\sigma_0^2}.
\end{equation}

\section{WDW equation with exponential potential}
\label{appWDWexppot}

The exponential potential is associated with a cosmological fluid with barotropic equation of state $p=w\rho$, $w=$ const.
The energy conservation equation in this case yields 
\begin{equation}
\rho=\rho_0 \left(\frac{a}{a_0}\right)^{3(1+w)}\quad \rightarrow \quad
\xi=x^{-3(1+w)},
\end{equation}
whereas the Friedmann equation leads to
\begin{equation}
\tau=\frac{2}{3(1+w)}x^{3(1+w)/2},
\end{equation}
with $w>-1$. 
The scalar field as well as the potential then turn out to be
\begin{equation}
\kappa(\phi_\pm-\phi_\pm^0)=\pm\sqrt{3(1+w)}\ln x , \qquad
\frac{V_\pm}{\rho_{\rm (crit),0}}=\frac12(1-w){\rm e}^{\mp\sqrt{3(1+w)}\kappa(\phi_\pm-\phi_\pm^0)},
\end{equation}
or equivalently
\begin{equation}
\kappa\phi_\pm=\pm\lambda\alpha, \qquad
V_\pm=V_0 {\rm e}^{\mp\lambda\kappa\phi_\pm},
\end{equation}
where $\alpha=\ln x$, $\lambda^2=3(1+w)$ and $V_0=(1-w)\rho_{\rm (crit),0}/2$, 
and we have set $\phi_\pm^0=0$ for simplicity.

\subsection{Solution in the BO approximation}

Let us consider the $V_+$ branch of the potential, so that the late time regime is reached for large positive values of $\phi=\phi_+$.
Substituting then the profile $V(\phi)=V_0 {\rm e}^{-\lambda\kappa\phi}$ into the WDW equation (\ref{(3.5)}) yields the following general solution for $\phi_k$ 
\begin{equation}
\phi_k(\alpha, \phi)=f_1(\alpha)\,I_{\nu}(x)+f_2(\alpha)\,K_{\nu}(x),
\end{equation}
where $I_{\nu}(x)$ and $K_{\nu}(x)$ are the modified Bessel functions of first and second kind, respectively, with
\begin{equation}
\nu = \frac{2\sqrt{-2E_k(\alpha)}}{\lambda\kappa\hbar},\qquad
x = \frac{2\sqrt{2V_0}}{\lambda\kappa\hbar}{\rm e}^{3\alpha-(\lambda\kappa/2)\phi},
\end{equation}
whereas $f_1(\alpha)$ and  $f_2(\alpha)$ are arbitrary functions of $\alpha$.
Regularity for large $\phi$ requires $f_1(\alpha)=0$, and one can set $f_2(\alpha)=1$ without any loss of generality.
The energy eigenvalues are
\begin{equation}
E_k(\alpha)=-\frac18\lambda^2\kappa^2\hbar^2k^2,
\end{equation}
so that the equation for the functions $C_k$ becomes
\begin{equation}
\left[{{\rm d}^{2}\over {\rm d}\alpha^{2}}-\frac14\lambda^2\kappa^2k^2\right] C_k=0,
\end{equation}
whose solution is 
\begin{equation}
C_k(\alpha)=c_1{\rm e}^{-(\lambda\kappa k/2)\alpha}+c_2{\rm e}^{(\lambda\kappa k/2)\alpha}.
\end{equation}
Regularity for large $\alpha$ requires $c_2=0$, and one can set $c_1=1$.
Therefore, the complete solution to the WDW equation is 
\begin{equation}
\label{BOsolexp}
\Psi(\alpha, \phi)= \sum_kN_k{\rm e}^{-(\lambda\kappa k/2)\alpha}\,K_{\nu}(x),
\end{equation}
where $N_k$ is a normalization factor.

\subsection{Exact solution}

The case of an exponential potential can also be solved exactly.
In fact, the WDW equation (\ref{wdweq}) reads
\begin{equation}
\label{wdweqexp}
\hbar^2\left[\partial_{\alpha\alpha}-\partial_{\phi\phi}\right]\Psi(\alpha, \phi)
+2{\rm e}^{6\alpha}V_0 {\rm e}^{-\lambda\kappa\phi}\Psi(\alpha, \phi)=0.
\end{equation}
If we change coordinates according to \cite{kiefer1}
\begin{eqnarray}
u& = &\frac{\sqrt{2V_0}}{3}\frac{{\rm e}^{Y}}{1-(\lambda\kappa/6)}
\left(\cosh X+\frac{\lambda\kappa}{6}\sinh X\right), 
\nonumber\\
v& = &\frac{\sqrt{2V_0}}{3}\frac{{\rm e}^{Y}}{1-(\lambda\kappa/6)^2}
\left(\sinh X+\frac{\lambda\kappa}{6}\cosh X\right),
\end{eqnarray}
where
\begin{equation}
X = 3\phi-(\lambda\kappa/2)\alpha,\qquad
Y = 3\alpha-(\lambda\kappa/2)\phi,
\end{equation}
Eq. (\ref{wdweqexp}) becomes
\begin{equation}
\label{wdweqexp_uv}
\hbar^2\left[\partial_{uu}-\partial_{vv}\right]\Psi(u, v) +\Psi(u, v)=0.
\end{equation}
This equation can be solved exactly, with general solution
\begin{equation}
\Psi(u, v)=\sum_kN_k{\rm e}^{k(u+v)-\frac1{4\hbar^2k}(u-v)}
=\sum_kN_k {\rm e}^{k_{u} u} {\rm e}^{k_{v} v} ,
\end{equation}
where $N_k$ is a normalization factor and
\begin{equation}
k_{u} = k  -\frac1{4\hbar^2k},\qquad
k_{v} = k  +\frac1{4\hbar^2k}.
\end{equation}

\subsection{Construction of wave packets}

However, one can use the JWKB approximation, i.e., interpreting the wave function in terms of classical 
trajectories according to the decomposition
\begin{equation}
\label{WKBsol}
\Psi(u, v)=\int {\rm d}k A_k\left[c_1{\rm e}^{({\rm i}/\hbar)S_k}+c_2{\rm e}^{-({\rm i}/\hbar)S_k}\right], 
\end{equation}
where each phase factor $S_k=S_k(u,v)$ is taken to obey the Hamilton-Jacobi equation $(\partial_uS_k)^2-(\partial_vS_k)^2=1$ at lowest order, 
with solution $S_k=ku-\sqrt{k^2-1}v$. This form of the solution is suitable to construct semiclassical wave packets and 
study their behavior close to the classical solution \cite{vilenkin,kiefer88}.
The latter is recovered through the principle of constructive interference, i.e., 
$\frac{\partial S_k}{\partial k}\vert_{k=\bar k}=0$, leading to $u/v={\bar k}/({\bar k}^2-1)$, where now $u$ and $v$, or equivalently $\alpha$ and $\phi$, are determined 
by the classical asymptotic SC dynamics. 
Choosing then for the amplitude a Gaussian with width $\sigma$ centered around $k=\bar k$ corresponding 
to the classical solution, i.e.,
\begin{equation}
A_k=\frac1{(\sqrt{\pi}\sigma\hbar)^{1/2}}{\rm e}^{-(k-\bar k)^2/2\sigma^2\hbar^2},
\end{equation}
one can study the spreading of wave packets around the classical trajectories.
Following Ref. \cite{kiefer1}, the wave packet finally reads
\begin{equation}
\label{Psiewbexp}
\Psi(u, v)=c_1\pi^{1/4}\sqrt{\frac{2\sigma\hbar}{1-{\rm i}\sigma^2\hbar S_0''}}
{\rm e}^{({\rm i}/\hbar)S_0-S_0'{}^2/2(\sigma^{-2}-{\rm i}\hbar S_0'')}+{\rm c.c.},
\end{equation}
where the phase factor $S_k$ has been Taylor expanded about $\bar k$ up to the second order (a prime 
denoting derivative with respect to $k$ and $S_0\equiv S_k(\bar k)$), c.c. indicates the complex conjugate of the preceding term, and we have set $c_1=c_2$ for simplicity.
The second derivatives $S_0''$ are a measure for the dispersion of the wave packet around the classical path.
In the case under consideration here, we have $S_0''=v/(\bar k^2-1)^{3/2}$, so that the width of the Gaussian increases without limit for large values of $v$ \cite{kiefer1}.

\subsection{The limiting case of constant potential}

The solution to the WDW equation with a constant potential $V(\phi)=V_0$ can be simply obtained from that corresponding to an exponential potential discussed above
in the limit $\lambda\to0$.
The WDW equation (\ref{wdweq}) then reads
\begin{equation}
\label{wdweqconst}
\hbar^2\left[\partial_{\alpha\alpha}-\partial_{\phi\phi}\right]\Psi(\alpha, \phi)
+2{\rm e}^{6\alpha} V_0\Psi(\alpha, \phi)=0,
\end{equation}
and can be cast in the same form as Eq. (\ref{wdweqexp_uv}) through the transformation
\begin{equation}
\alpha \equiv \frac16\ln\left[\frac{9}{2V_0}(u+v)(u-v)\right],\qquad
\phi \equiv \frac16\ln\left(\frac{u+v}{u-v}\right),
\end{equation}
with inverse 
\begin{equation}
u=\frac{\sqrt{2V_0}}{3}{\rm e}^{3\alpha}\cosh 3\phi, \qquad
v=\frac{\sqrt{2V_0}}{3}{\rm e}^{3\alpha}\sinh 3\phi.
\end{equation}
The JWKB solution is still given by Eq. (\ref{WKBsol}).

\begin{acknowledgements}
D. B. acknowledges ICRANet for partial support. G. E. is grateful to the Dipartimento di Fisica Ettore Pancini of Federico II University, Naples, for hospitality and support. The authors are indebted 
with Profs. C. Kiefer and S. Succi for useful discussions.
\end{acknowledgements}

\end{document}